\documentclass[10pt]{article}

\usepackage{amssymb}
\usepackage{amsfonts}
\usepackage{amsmath}
\usepackage{amscd}
\usepackage[all,cmtip]{xy}
\usepackage{amsthm}
\usepackage{setspace}
\usepackage{enumerate}
\usepackage{graphicx}
\usepackage{float}
\usepackage{tikz}
\usepackage{hyperref}

\usepackage{geometry}

\usetikzlibrary{decorations.pathmorphing}
\usetikzlibrary{patterns}
\usetikzlibrary{shapes.geometric}
\usetikzlibrary{cd}

\theoremstyle{plain}

\newtheorem*{example}{Example}
\newtheorem{thm}{Theorem}

\theoremstyle{definition}

\newtheorem{defi}{Definition}

\renewcommand{\AA}{{\mathbb A}}
\newcommand{\CC}{{\mathbb C}}

\newcommand{\ZZ}{{\mathbb Z}}

\newcommand{\ol}{\overline}

\newcommand{\h}{{\hbar}}

\title{Langlands parameters of quivers in the Sato Grassmannian}
\author{Martin T. Luu, Matej Penciak}

\date{}

\begin{document}

\maketitle

\newcommand{\Addresses}{{
\bigskip
\footnotesize

M. Luu, \textsc{Department of Mathematics, University of Illinois at Urbana-Champaign, IL 61801, USA} \par \nopagebreak \textit{E-mail address:} \texttt{mluu@illinois.edu}

\medskip

M. Penciak, \textsc{Department of Mathematics, University of Illinois at Urbana-Champaign, IL 61801, USA} \par \nopagebreak \textit{E-mail address:} \texttt{penciak2@illinois.edu}

}}

\begin{abstract} 
Motivated by quantum field theoretic partition functions that can be expressed as products of tau functions of the KP hierarchy we attach several types of local geometric Langlands parameters to quivers in the Sato Grassmannian. We study related questions of Virasoro constraints, of moduli spaces of relevant quivers, and of classical limits of the Langlands parameters. 
\end{abstract}

\section{Introduction}
\label{introduction}
A Langlands parameter for the local field $\mathbb{F}_{q}(\!(t)\!)$ consists of a finite dimensional Weil-Deligne representation of 
the Weil group of $\mathbb{F}_{q}(\!
(t)\!)$. The local arithmetic Langlands correspondence relates such parameters to representations of $\textrm{GL}_{n}(\mathbb{F}_{q}(\!(t)\!))$. In the local geometric Langlands correspondence for the general linear group the role of Langlands parameters is played by connections on the formal punctured disc $\textrm{Spec } \CC(\!(t)\!)$. In the proposed local geometric Langlands correspondence of Frenkel-Gaitsgory \cite{FG} such parameters are related to categorical representations of $\textrm{GL}_{n}(\CC(\!(t)\!))$. A crucial input in the construction of this correspondence is played by the results of Feigin-Frenkel on the center of vertex algebras associated to affine Lie algebras. In this way quantum field theory plays a role in the geometric correspondence. 

In the present work we aim to develop further a different relation between quantum field theories and connections on the punctured disc. The intermediate objects are special cases of what we call quivers in the Sato Grassmannian. The point is that some quantum field theory partition functions can be expressed in terms of the $\tau$-functions associated to vertices of such quivers. Furthermore, in some situations the quiver gives rise to various connections on the formal punctured disc and in this manner one obtains the relation between some local geometric Langlands parameters and quantum field theory. Our work is motivated by the attempt to describe a natural D-module theoretic framework in the context of the philosophy of Kac-Schwarz operators that is wide enough to deal for example with the work by Kac-Schwarz \cite{KS} on minimal models coupled to gravity as well as the work by Anagnostopoulos-Bowick-Schwarz \cite{ABS} on the string equation for unitary matrix models.

Fix an indeterminate $z$ and let $Gr$ denote the big cell of the index zero part of the Sato Grassmannian: It consists of complex subspaces of $\CC(\!(1/z)\!)$ whose projection to $\CC[z]$ is an isomorphism. 
\begin{defi}
Fix an integer $n \ge 1$. A quiver in the Sato Grassmannian denotes a collection of operators
$$U_{k} \in \CC(\!(1/z)\!)[\partial_{z}], \;\;\; k \ge 1$$
together with the data for each $k$ of a subset $T_{k} \subseteq \{1,2,\cdots,n\}$ and a map 
$$s_{k} : T_{k} \rightarrow \{1,2,\cdots, n\}.$$ 
A solution to the quiver is defined to be an $n$-tuple $(V_{1},\cdots, V_{n})$ of points of the big cell of the Sato Grassmannian such that 
$$U_{k} V_{i} \subseteq V_{s_{k}(i)}$$
for all $i \in T_{k}$. We call $n$ the degree of the quiver. 
\end{defi}
One can describe quivers and their solutions by letting the points $V_{i}$ correspond to vertices of a graph and letting the operators $U_{k}$ correspond to directed edges. 

We now give two examples of quivers that are related to quantum field theory. The first one concerns the symmetric unitary matrix model and the motivation in physics comes for example from describing two dimensional QCD with unitary gauge group. 

Fix a positive integer $N$ and a polynomial potential
$$V(X)=\sum_{i \ge 0} g_{i} X^{i}.$$
The partition function of the symmetric unitary one-matrix model is defined as
$$Z_{N} =\int_{U(n)} \exp \left( - \frac{N \cdot \textrm{Tr } V(X +X^{\dagger})}{\lambda} \right ) \textrm{ d}\mu$$
where $\textrm{d}\mu$ is a Haar measure on the group $U(n)$ of $n\times n$ unitary matrices and $N=[n/2]$ and $(-)^{\dagger}$ denotes the conjugate transpose.
Let $Z$ denote the partition function of the double scaling limit. 
Consider a function $v$ that satisfies
$$v^{2} = - \partial^{2}_{x} \ln Z$$
where $x$ is the sum of the two variables that are held fixed in the double scaling limit. The function $v$
is a solution of the modified KdV (mKdV) equation. For a description of $Z$ in terms of the Sato Grassmannian it is useful to relate $v$ via the Miura transform to the usual KdV equation: $u_{1/2} = v^{2} \pm v_{x}$ gives a solution $u_{1/2}$ of the KdV equation. Then
$$u_{1/2}= - 2 \partial^{2}_{x} \ln \tau_{1/2}$$
where $\tau_{1/2}$ is a $\tau$-function of the KdV hierarchy. It is known, see for example the work of Anagnostopoulos-Bowick-Schwarz \cite{ABS}, that the partition function satisfies
$$Z = \tau_{1} \cdot \tau_{2}$$
where $\tau_{1}$ and $\tau_{2}$ are KdV $\tau$-functions whose associated points $V_{1}$ and $V_{2}$ of the Sato Grassmannian satisfy certain constraints. Namely
\begin{enumerate}[(i)]
\item
$zV_{1}\subseteq V_{2}$ 
\item
$z V_{2} \subseteq V_{1}$
\item
$AV_{1}\subseteq V_{2}$
\item
$ A V_{2} \subseteq V_{1}$
\end{enumerate}
where $A$ is an operator depending on the potential of the matrix model. A simple example is
$$A=  \frac{\textrm{d}}{\textrm{d} z} + z^{2n}$$
for some $n \ge 1$. One obtains a quiver in the Sato Grassmannian that can be described in the following manner:
$$
\begin{tikzpicture}
\draw[->] (1.5,4.3) arc (-45 : -135 : -1.4 cm) ;
\draw[->] (1.7,4.1) arc (-80 : -100 : -4.5 cm) ;
\draw (2.5,5.2) node {$A$};
\draw (2.5,4.3) node {$z$};
\draw (1.5,4.0) node[circle, inner sep=1.5pt,fill] {} ;
\draw (3.5,4.0) node[circle, inner sep=1.5pt,fill] {} ;
\draw[->] (3.5,3.7) arc (-45 : -135 : 1.4 cm) ;
\draw[->] (3.3,3.9) arc (-80 : -100 : 4.5 cm) ;
\draw (2.5,2.8) node {$A$};
\draw (2.5,3.6) node {$z$};
\end{tikzpicture}
$$
Another important example of quivers related to quantum field theory concerns the $(q,p)$ minimal model conformal field theory coupled to gravity. Here $q$ and $p$ are positive co-prime integers. The corresponding partition function can be described as a square of a special KP $\tau$-function satisfying two additional constraints. In terms of the Sato Grassmannian the problem reduces to finding a point $V \in Gr$ stabilized by $z^{q}$ and the Kac-Schwarz operator $A^{q,p}$: The situation is described by a quiver of the form
$$
\begin{tikzpicture}
\draw[->] (2.7,4.2) arc (-70 : 250 : 0.6 cm) ;
\draw (4,4.8) node {$A^{q,p}$};
\draw (2.5,4.0) node[circle, inner sep=1.5pt,fill] {} ;
\draw[->] (2.3,3.8) arc (-250 : 70 : 0.6 cm) ;
\draw (4,3.2) node {$z^{q}$};
\end{tikzpicture}
$$
where
$$A^{q,p} =  \partial_{z^{q}}+ \frac{1-q}{2q} \frac{1}{z^{q}}+z^{p}$$
and
$$\partial_{z^{q}} = \frac{1}{qz^{q-1}} \frac{\textrm{d}}{\textrm{d} z}.
$$

\section{String quivers}
An important special case of our considerations is related to operators $P$ and $Q$ in $(\mathfrak g \mathfrak l_{n} \CC[\![x]\!] )[\partial_{x}]$ such that
$$[P,Q]=1.$$
The importance of such differential operators in describing partitions functions of quantum field theories has long been recognized, the case of topological gravity is a well known example. Such $P$ and $Q$ yield a homomorphism from the one variable Weyl algebra. The case where the operator $P$ is monic is studied in the work of Schwarz \cite{SCH} on quantum curves.

We now define special types of quivers that are the focus of the present work. They include the two types of examples from the previous section and are motivated by the philosophy of Kac-Schwarz operators \cite{KS}. The starting point are two operators
$$\mathcal P, \mathcal Q \in \CC(\!(1/z)\!)[\partial_{z}]$$
such that
$$[\mathcal P,\mathcal Q]=1.$$
We consider quivers where the operators $U_{k}$ are all either $\mathcal P$ or $\mathcal Q$. We specialize from now on to the case where 
$$\mathcal Q =z^{q}$$ 
for some integer $q \ge 1$ and
$$\mathcal P = \partial_{z^{q}} + \sum_{i\le p} c_{i} z^{i}$$
for some integer $p\ge 1$. Let $n$ denote the degree of the quiver. We assume that
$$\mathcal P V_{i} \subseteq V_{\sigma_{\mathcal P}(i)}$$
$$\mathcal Q V_{i} \subseteq V_{\sigma_{\mathcal Q}(i)}$$
where $\sigma_{\mathcal P}$ and $\sigma_{\mathcal Q}$ are $n$-cycles and we view them as maps $\ZZ/n\ZZ \rightarrow \ZZ/n\ZZ$. We make the convention that 
$$\sigma_{\mathcal Q}(i) = i+1.$$ 
Suppose $(V_{1},\cdots,V_{n})$ is a solution to the quiver and suppose $\phi_{i}$ is in $V_{i}$. One has
$$\mathcal P \mathcal Q \phi_{i} = \mathcal Q \mathcal P \phi_{i} + \phi_{i}.$$
The three terms are elements of $V_{\sigma_{\mathcal P}(i+1)}$, $V_{\sigma_{\mathcal P}(i)+1}$, and $V_{i}$. It follows that the study of this type of quiver simplifies if 
$$\sigma_{\mathcal P}(i)=i-1.$$
In this case the permutations $\sigma_{\mathcal P}$ and $\sigma_{\mathcal Q}$ are inverses to each other. This yields the following notion: 

A solution to a degree $n$ string quiver is an $n$-tuple $(V_{1},\cdots,V_{n})$ of points in $Gr$ such that
$$z^{q}V_{i} \subseteq V_{i+1}$$
and
$$\mathcal P V_{i} := \left (\partial_{z^{q}} + f(z) \right ) V_{i} \subseteq V_{i-1}$$
for all $i$, where the indices are taken modulo $n$. We denote this quiver by
$$\textrm{String}(n,q,f).$$ 
We prove various results on D-modules related to string quivers and some of them carry over, in modified form, to non string quivers, see Section \ref{non-string-section}. A slight perturbation of the quiver can have strong effects on the collection of solutions. For example, a degree $7$ string quiver is depicted in Figure \ref{second-figure} and a description of the solution space in many cases follows from Theorem \ref{moduli-theorem}. On the other hand, as will be discussed in Section \ref{non-string-section}, the solution space of the non string quiver in Figure \ref{non-string-figure} can be much simpler.

\begin{figure}[h!]
\centering
\begin{tikzpicture}

\node (1) at ({360/7*0}:60pt) {$\bullet$};
\node (2) at ({360/7*1}:60pt) {$\bullet$};
\node (3) at ({360/7*2}:60pt) {$\bullet$};
\node (4) at ({360/7*3}:60pt) {$\bullet$};
\node (5) at ({360/7*4}:60pt) {$\bullet$};
\node (6) at ({360/7*5}:60pt) {$\bullet$};
\node(7) at ({360/7*6}:60pt) {$\bullet$};

\path[commutative diagrams/.cd, every arrow, every label]
	(1) edge[bend right] node[swap] {$\mathcal Q$} (2)
	(2) edge[bend right] node[swap] {$\mathcal Q$} (3)
	(3) edge[bend right] node[swap] {$\mathcal Q$} (4)
	(4) edge[bend right] node[swap] {$\mathcal Q$} (5)
	(5) edge[bend right] node[swap] {$\mathcal Q$} (6)
	(6) edge[bend right] node[swap] {$\mathcal Q$} (7)
	(7) edge[bend right] node[swap] {$\mathcal Q$} (1)
	(1) edge node[swap] {$\mathcal P$} (7)
	(7) edge node[swap] {$\mathcal P$} (6)
	(6) edge node[swap] {$\mathcal P$} (5)
	(5) edge node[swap] {$\mathcal P$} (4)
	(4) edge node[swap] {$\mathcal P$} (3)
	(3) edge node[swap] {$\mathcal P$} (2)
    (2) edge node[swap] {$\mathcal P$} (1)	;

\end{tikzpicture}
\caption{Degree $7$ string quiver}
\label{second-figure}
\end{figure}

\begin{figure}[h!]
\centering
\begin{tikzpicture}

\node (1) at ({360/7*0}:60pt) {$\bullet$};
\node (2) at ({360/7*1}:60pt) {$\bullet$};
\node (3) at ({360/7*2}:60pt) {$\bullet$};
\node (4) at ({360/7*3}:60pt) {$\bullet$};
\node (5) at ({360/7*4}:60pt) {$\bullet$};
\node (6) at ({360/7*5}:60pt) {$\bullet$};
\node(7) at ({360/7*6}:60pt) {$\bullet$};

\path[commutative diagrams/.cd, every arrow, every label]
	(1) edge[bend right] node[swap] {$\mathcal Q$} (2)
	(2) edge[bend right] node[swap] {$\mathcal Q$} (3)
	(3) edge[bend right] node[swap] {$\mathcal Q$} (4)
	(4) edge[bend right] node[swap] {$\mathcal Q$} (5)
	(5) edge[bend right] node[swap] {$\mathcal Q$} (6)
	(6) edge[bend right] node[swap] {$\mathcal Q$} (7)
	(7) edge[bend right] node[swap] {$\mathcal Q$} (1)
	(1) edge node[swap] {$\mathcal P$} (3)
	(2) edge node[swap] {$\mathcal P$} (4)
	(3) edge node[swap] {$\mathcal P$} (5)
	(4) edge node[swap] {$\mathcal P$} (6)
	(5) edge node[swap] {$\mathcal P$} (7)
	(6) edge node[swap] {$\mathcal P$} (1)
    (7) edge node[swap] {$\mathcal P$} (2)	;

\end{tikzpicture}
\caption{Degree $7$ non string quiver}
\label{non-string-figure}
\end{figure}

\subsection{D-modules for string quivers}
\label{D-module-section}
Fix a coordinate $t$ to describe the formal punctured disc as
$$D^{\times} = \textrm{Spec }\CC(\!(t)\!)$$ 
and let
$$u:=\frac{1}{t}.$$ 
We now define and calculate various D-modules attached to string quivers. These D-modules are connections on the formal punctured disc $\textrm{Spec }\CC(\!(t)\!)$. Such an object consists of a $\CC(\!(t)\!)$-vector space $V$ together with a $\CC$-linear endomorphism $\nabla$ of $V$ such that for all $f \in \CC(\!(t)\!)$ and $v \in V$ one has
$$\nabla(fv) = (\partial_{t}f)v + f \nabla(v).$$ 
The collection of such connections forms a category where morphisms between $(V_{1},\nabla_{1})$ and $(V_{2},\nabla_{2})$ are $\CC(\!(t)\!)$-linear maps $g : V_{1} \rightarrow V_{2}$ such that
$$g \circ \nabla_{1} = \nabla_{2} \circ g.$$ 
Given a $\CC[u,\partial_{u}]$-module $M$ one obtains a D-module on $\AA^{1}= \textrm{Spec } \CC[u]$. The restriction to the formal punctured disc $D^{\times}$ around $\infty$ is defined as
$$\textrm{Res}_{\infty}(M):= M \otimes_{\CC[u]} \CC(\!(t)\!).$$
It naturally has the structure of connection on the formal puncture disc $\textrm{Spec } \CC(\!(t)\!)$. 

Some notation: Given a one dimensional connection $\nabla=\partial_{t} + \phi$ on $D^{\times}$ and an integer $q \ge 1$ we denote by $[q]_{*}\nabla$ the push-forward under the map $t \mapsto t^{q}$ of $\partial_{t}+qt^{q-1}\phi$. We also write connections on $D^{\times}$ in terms of the variable $u=1/t$. So for example, by $(\CC(\!(t)\!),\partial_{u}+1)$ we denote the $1$-dimensional irregular connection 
$(\CC(\!(t)\!),\partial_{t}- 1/t^{2})$. 

\subsubsection{D-modules I}
Since $[\mathcal P,\mathcal Q]=1$ one obtains an action of the Weyl algebra $\CC[u,\partial_{u}]$ by letting $u\mapsto \mathcal Q$ and $\partial_{u}\mapsto \mathcal P$ act on $\CC(\!(1/z)\!)$. By restricting to the formal punctured disc around $\infty$ one obtains the structure of a connection $\nabla_{\textrm{I}}$ on $D^{\times} = \textrm{Spec }\CC(\!(t)\!)$. In fact
 $$\nabla_{\textrm{I}}\cong [q]_{*}\left( \CC(\!(t)\!), \partial_{u} + f(u) \right ).$$
 
\subsubsection{D-modules II}
\label{string-section}
Let $n\ge 1$ and let $Gr^{n}$ denote the $n$-fold product of the big cell of the Sato Grassmannian. Suppose given a solution $(V_{1},\cdots,V_{n})$ to $\textrm{String}(n,q,f)$. Define the $\CC$-vector space
$$M:= V_{1} \times \cdots \times V_{n}.$$
Let $\tilde  P \in \textrm{End}_{\CC}(M)$ be defined via
$$\tilde P (v_{1},\cdots,v_{n}) = (\mathcal P v_{\sigma^{-1}_{\mathcal P}(1)}, \cdots, \mathcal P v_{\sigma^{-1}_{\mathcal P}(n)})$$
and let $\tilde Q \in \textrm{End}_{\CC}(M)$ be defined via
$$\tilde Q (v_{1},\cdots,v_{n}) = (\mathcal Q v_{\sigma^{-1}_{\mathcal Q}(1)}, \cdots, \mathcal Q v_{\sigma^{-1}_{\mathcal Q}(n)}).$$
Then
$$
\tilde P \tilde Q (v_{1},\cdots,v_{n}) - \tilde Q \tilde P(v_{1},\cdots v_{n}) = ( \cdots,\mathcal P \mathcal Q v_{i} -\mathcal Q \mathcal P v_{i},\cdots ) $$
and since
$[\mathcal P,\mathcal Q]=1$ it follows that
$$[\tilde P,\tilde Q]= 1.$$
As a consequence, the $\CC$-vector space $M$ carries a representations of the one-variable Weyl algebra $\CC[u,\partial_{u}]$ by letting $\partial_{u}$ act via $\tilde P$ and $u$ via $\tilde Q$. 

\begin{defi}
\label{Kac-Schwarz-definition}
The connection $\nabla_{\textrm{II}}$ of the quiver $\textrm{String}(n,q,f)$  is defined to be 
$$\nabla_{\textrm{II}}:=\textrm{Res}_{\infty}(M).$$ 
\end{defi}

\subsubsection{D-modules III}
Consider the quiver $\textrm{String}(n,q,f)$ and suppose $V=(V_{1},\cdots,V_{n})$ is a solution to the quiver. Let 
$$\pi : \CC(\!(1/z)\!) \rightarrow \CC[z]$$
be the projection map to the polynomial part of a Laurent series with respect to $1/z$. For a point $V$ in the big cell of the Sato Grassmannian there is a unique element in the intersection of $V$ and $\pi^{-1}(z^{s})$ for $s \ge 0$, we denote it by $\pi^{-1}(z^{s})_{V}$. Furthermore, we use the notation $a \sim b$ if $a$ is an object with an expression 
$$a= b + \textrm{lower order terms with respect to $z$}.$$
Let
$$\phi_{r,s} =\pi^{-1}(z^{s})_{V_{r}}$$
for $1\le r \le n$ and $0 \le s \le q-1$. Let $\Phi_{r,s}$ be the element of $M:=V_{1}\times \cdots \times V_{n}$ with all zero entries except that the $r$'th component is given by $\phi_{r,s}$. We now show that one obtains a basis for $M$ as a $\CC[\tilde Q]$-module. For each $i$ let $\iota V_{i}$ denote the image of $V_{i}$ in $M$ under the natural inclusion map. For $k=(r-1)q+s$ with $0\le s <q$ let
$$\xi_{k}= \phi_{r,s} \;\;\; \textrm{ and } \;\;\; \Xi_{k} = \Phi_{r,s}.$$
One has 
$$z^{s+kq}\sim \tilde Q^{k} \Phi_{i-k,s} \in \iota V_{i} $$
and
$$\textrm{span}_{\CC} \left \{\tilde Q^{k} \Phi_{i-k,s} \; | \; 0\le s\le q-1 \textrm{ and }  k\ge 0 \right  \} = \iota V_{i}$$
where $i-k$ is considered modulo $n$, and hence one has a spanning set. On the other hand, any polynomial in $\tilde Q$ raises the $z$ degree of every non-zero element of $M$ and hence $M$ is a torsion-free and hence free $\CC[\tilde Q]$-module, and in fact the $\Phi_{r,s}$'s yield a basis. Define $B_{l,k}$ via 
$$\tilde P \Xi_{k} = \sum_{0\le l < nq} B_{l,k}(\tilde Q) \Xi_{l}$$
and let 
$$B = (B_{l,k}(z^{q}))=(B_{l,k}(\mathcal Q)) \in \mathfrak g \mathfrak l_{nq} ( \CC[z^{q}] ).$$

\begin{defi}
\label{companion-connection-definition}
The connection $\nabla_{\textrm{III}}^{\pm}$ of the quiver $\textrm{String}(n,q,f)$ is defined to be the restriction to the punctured disc around $\infty$ of the connection $\partial_{u} \pm B(u)$.
\end{defi}
\begin{defi}
Consider a permutation $\sigma \in S_{n}$ and denote also by $\sigma$ the corresponding map $\ZZ/n\ZZ \rightarrow \ZZ/n\ZZ$. Define the subset 
$$\mathcal B(\sigma) \subseteq \mathfrak g \mathfrak l_{nq} ( \CC[z^{q}] )$$ to consist of matrices $B=(B_{l,k})$ satisfying 
$$B_{l,k} = \sum_{t \ge 0} b_{l,k,t} z^{qt}$$ 
such that
$b_{l,k,t} = 0$ except possibly for 
$$t\equiv  \sigma ( \lfloor \frac{k}{q} \rfloor +1) - (\lfloor \frac{l}{q} \rfloor +1) \mod n.$$
\end{defi}
We now show:
\begin{thm}
\label{companion-theorem}
Consider the quiver $\textrm{\emph{String}}(n,1,f)$ such that 
$\textrm{\emph{gcd}}(1+\textrm{\emph{deg }} f, n)=1$. The gauge equivalence class of $\nabla_{\textrm{\emph{III}}}^{+}$ is independent of the choice of solution and in fact if $\zeta_{n}$ is a primitive $n$'th root of unity one has
$$ \nabla_{\textrm{\emph{II}}}  \cong \nabla_{\textrm{\emph{III}}}^{+}\cong \bigoplus_{i=0}^{n-1} \left (\CC(\!(t)\!),\partial_{u} +\zeta_{n}^{i} f(\zeta_{n}^{i}u) \right ).$$
\end{thm}
\begin{proof}
For $B$ be as before one sees that $B$ describes the connection $\nabla_{\textrm{II}}$ in the basis $\Phi_{r,s}$ and it follows that $\nabla_{\textrm{II}} \cong \nabla_{\textrm{III}}^{+}$. We now show that $B$ is in $\mathcal B(\sigma_{\mathcal P})$. Consider $k$
with $0\le k <n$. Then 
$$\Xi_{k} \in \iota V_{k +1}.$$ 
Note that $\tilde Q^{i}$ takes $\Xi_{l}$ to an element in $\iota V_{ l+1+i}$ and $\tilde P$ takes $\Xi_{k}$ to an element of $\iota V_{k}$. Hence if $b_{l,k,i} \ne 0$ one needs
$$k \equiv l +1+i  \mod n$$
and hence
$$i \equiv \sigma_{\mathcal P} ( k +1) - (l +1) \mod n.$$
It follows that 
$$B \in \mathcal B(\sigma_{\mathcal P}).$$
Now note that
$$(\partial_{z}+f) \begin{bmatrix}
\xi_{1}\\
\vdots \\
\xi_{n}
\end{bmatrix}=B\begin{bmatrix}
\xi_{1}\\
\vdots \\
\xi_{n}
\end{bmatrix}.$$
It follows that $-f$ is an exponential factor of $\nabla_{\textrm{III}}^{-}$. We claim that the leading order coefficient matrix of $B$ is diagonalizable with $n$ distinct eigenvalues. From this it follows that $f$ is an exponential factor of $\nabla_{\textrm{III}}^{+}$. Let $D$ be the maximum degree of $B$ with respect to $z$. One has
$$\tilde P \Xi_{k} \sim z^{\textrm{deg }f}$$
and since $q=1$ one has for all $a,b$ that
$$\tilde Q^{k} \Phi_{a,b} \sim z^{k}.$$
Since $B \in \mathcal B(\sigma_{\mathcal P})$ it follows that $b_{l,k,t}=0$ unless possibly if
$$l\equiv k - t-1 \mod n$$ 
and hence if $b_{l,k,t} \ne 0$ then $t \le \textrm{deg }f$ and one can deduce
$$D  =  \textrm{deg }f.$$
Furthermore, for all $k,l$ such that $l\equiv k - D-1 \mod n$ the value $b_{l,k, D}$ has the same non-zero value. Hence, there is $r \in \CC^{\times}$ such that the leading order coefficient matrix $\tilde B$ of $B$ is $r$ times the permutation matrix corresponding to
$$\sigma : k \mapsto k - (\textrm{deg }f +1) \mod n.$$
and since we assume that $\textrm{gcd}(1+\textrm{deg }f,n)=1$, it follows that $\sigma$ is a $n$-cycle and the desired result follows.

Suppose now $(V_{1},\cdots,V_{n})$ is a solution to the quiver. Let $\zeta$ be an $n$'th root of unity and let $d$ denote the diagonal $n \times n$ matrix whose $k$'th diagonal entry is given by $\zeta^{k}$. Then the $(b,a)$ entry of
$$d\zeta B(\zeta z) d^{-1}$$
equals
$$\zeta^{- a+b} \cdot \zeta \cdot \zeta^{a - b -1} B(z)_{b,a} = B(z)_{b,a}$$
since $B \in \mathcal B(\sigma_{\mathcal P})$. Therefore, if $f$ is an exponential factor of $\nabla_{\textrm{II}}$ then for all $h\ge 1$
$$\zeta^{h} f(\zeta^{h}z)$$
is an exponential factor as well. Hence for each $0 \le i <n$ the connection
$$ \left (\CC(\!(t)\!),\partial_{u} +\zeta^{i} f(\zeta^{i}u) \right  )$$
is isomorphic to a sub-connection of $\nabla_{\textrm{II}}$. To complete the determination of the Levelt-Turrittin normal form of $\nabla_{\textrm{II}}$ it suffices to show that the above connections  are non-isomorphic for distinct choices of $i$. Hence, suppose that $0 \le i_{0},i_{1} < n$ and 
$$\zeta^{i_{0}}f(\zeta^{i_{0}} w) - \zeta^{i_{1}}f(\zeta^{i_{1}} w) \in \frac{1}{w} \CC[\![\frac{1}{w}]\!]. $$
Looking at the leading order coefficient it follows that 
$$i_{0}(1 +\textrm{deg } f) \equiv i_{1} (1+ \textrm{deg }f ) \mod n.$$
Hence, since $\textrm{gcd}(1+\textrm{deg } f, n)=1$ it follows that $i_{0} \equiv i_{1} \mod n$ and the theorem follows from the Levelt-Turrittin classification.
\end{proof}
Note in particular that $\nabla_{\textrm{I}}$ is always a direct summand of $\nabla_{\textrm{II}}$.

Another consequence of the theorem: The quiver attached to the symmetric unitary matrix model that is deformed via sources $t_{1}, t_{2}, \cdots$ from the multicritical points satisfies
$$\nabla_{\textrm{II}} \cong \left(\CC(\!(t)\!), \partial_{u} - 
\sum (2i+1)t_{2i+1}u^{2i} \right ) \oplus \left(\CC(\!(t)\!), 
\partial_{u} + \sum (2i+1)t_{2i+1}u^{2i} \right ).$$
The reason is that it is shown in \cite{ABS} that the coefficients $a_{i}$ of $f$ are related to the deformation parameters $t_{i}$ via $a_{i}=0$ for odd $i$ and
$$a_{2i}=-(2i+1)t_{2i+1}.$$
Hence the result follows from Theorem \ref{companion-theorem}.

\subsection{Description of the moduli space}
We now give an analogue for string quivers of arbitrary degree of the description in \cite{ABS} (Section 5) of the moduli space of solutions of the string equation of the symmetric unitary matrix model. 

To state the theorem in a precise manner we introduce some notions. Let $\mathcal B(\sigma,s)$ denote the subset of $\mathcal B(\sigma)$ consisting of matrices that are of degree $s$ with respect to $z^{q}$ and such that all non-zero entries of the leading order coefficient matrix with respect to $z^{q}$ have the same value. We will use this notion in the case where $\sigma$ is given by the permutation $\sigma_{\mathcal P}$ that takes $i$ in $\ZZ/n\ZZ$ to $i-1$. 

Theorem \ref{moduli-theorem} will specialize to the case $q=1$. Suppose $(V_{1},\cdots,V_{n})$ is a solution to the string quiver $\textrm{String}(n,1,f)$. For every
$$\gamma=1+\sum_{i \ge 1} r_{i} z^{-i}$$ 
with constant coefficients $r_{i}$ one has that $(\gamma V_{1}, \cdots, \gamma V_{n})$ is a solution to $\textrm{String}(n,1,\tilde f)$ for a suitable $\tilde f$. We call solutions that are related in this manner gauge equivalent solutions.

\begin{thm}
\label{moduli-theorem}
The moduli space $\mathcal S_{n}$ of gauge equivalence classes of solutions to the string quiver $\textrm{\emph{String}}(n,1,f)$ with varying $f$ such that
$\textrm{\emph{gcd}}(1+\textrm{\emph{deg }} f, n)=1$
is an $n$-fold unbranched cover of the space of matrices 
$$\bigsqcup_{\substack{ \textrm{\emph{gcd}}(1+s,n)=1 \\  \\ s \ge 1}} \mathcal B(\sigma_{\mathcal P},s).$$
\end{thm} 
\begin{proof}
For a solution $(V_{1},\cdots,V_{n})$ to the quiver we attach $B$ as in Definition \ref{companion-connection-definition}. We have already shown that $B \in \mathcal B(\sigma_{\mathcal P})$ and furthermore,  the maximum power $D$ of $\mathcal Q=z$ occurring in $B$ equals $\textrm{deg }f$ and to every solution to the quiver one can then attached an element of 
$$\bigsqcup_{\substack{ \textrm{gcd}(1+s,n)=1 \\  \\ s \ge 1}} \mathcal B(\sigma_{\mathcal P},s).$$ 
This element is independent of the choice of gauge equivalence class of solutions.

We now show that conversely, starting with an element $B$ in this space, one can construct solutions to the quiver. Let $-f$ be an exponential factor of $\nabla_{\textrm{III}}^{-}$. Then there is $\textbf{v}$ such that
$$\nabla_{\textrm{III}}^{-}\textbf{v} = (\partial_{z}-B) \textbf{v}=\overline{0}$$
with
$$\textbf{v}= \exp ( \int f(z) \textrm{ d}z  ) [\phi_{1}, \cdots,\phi_{n}]^{\textrm{T}}$$
where the $\phi_{i}$'s are elements of $\CC(\!(1/z)\!)$. For $\ol{\phi}:=[\phi_{1},\cdots,\phi_{n}]^{\textrm{T}}$ one obtains
$$(\partial_{z} + f(z)) \overline{\phi} = B \overline{\phi}.$$
Consider as before the $n$-cycle given by
$$\sigma : k \mapsto k - (D +1) \mod n.$$
For a primitive $n$'th root of unity $\zeta_{n}$ let $\beta_{i}$'s be defined via
$$\phi_{\sigma^{-s}(1)}:=  \exp( -\int f(z) \textrm{ d}z) \sum_{k=1}^{n} \zeta_{n}^{sk} \beta_{k}.$$
and let $\ol{\beta}:=[\beta_{1},\cdots,\beta_{n}]^{\textrm{T}}$. Then there is a matrix $C$ with
$\partial_{z} \ol{\beta} = C \ol{\beta}$ and such that the leading order coefficient matrix 
of $C$ is diagonal with $n$ distinct eigenvalues 
$$r\cdot \zeta_{n}^{i} $$ with 
$1\le i\le n$ and $r \in \CC^{\times}$ and note also that then $f(z)\in \CC(\!(1/z)\!)$. It follows from the statement about distinctness of eigenvalues that the connection $\partial_{z}-C$ 
can be diagonalized by a matrix of the form $
\sim \textrm{id}_{n}$ and therefore, after possibly scaling, one can assume
$$ \exp( -\int f(z) \textrm{ d}z) \beta_{i} \sim \delta_{i-i_{0},0}$$
for some $i_{0}$ only depending on $\sigma$. It follows that after possibly scaling all the $\phi_{i}$'s by the same scalar one has
$$\phi_{\sigma^{-s}(1)} =   \exp( -\int f(z) \textrm{ d}z) \sum_{k=1}^{n} \zeta_{n}^{sk} \beta_{k} \sim \zeta_{n}^{si_{0}}  \in \CC^{\times}$$
for all $1 \le s \le n$. Hence
$$V_{i}:= \textrm{span}_{\CC}(\phi_{i},z\phi_{i-1},z^{2} \phi_{i-2}, \cdots)$$ 
gives a point of the big cell $Gr$ of the Sato Grassmannian. We claim that $(V_{1},\cdots,V_{n})$ is a solution to the quiver $\textrm{String}(n,1,f)$ where $f$ is as above. Due to our definition of the $V_{i}$'s it follows that
$$zV_{i} \subseteq V_{i+1}.$$
Furthermore, for $\mathcal P := \partial_{z}+f$ one has for each integer $k \ge 0$
\begin{eqnarray*}
\mathcal P z^{k}\phi_{i}=z^{k} \mathcal P \phi_{i} + kz^{k-1}\phi_{i}&=&z^{k} \sum_{j=1}^{n}B_{j,i} \phi_{j} + kz^{k-1}\phi_{i}\\
&=&z^{k} \sum_{j=1}^{n} \sum_{t} b_{j,i,t} z^{t} \phi_{j} + kz^{k-1}\phi_{i} .
\end{eqnarray*}
Since $B \in \mathcal B(\sigma_{\mathcal P})$ it follows that $b_{j,i,t}=0$ unless possibly if $t\equiv i-j-1 \mod n$. It follows that 
$$z^{k} b_{j,i,t}z^{t} \phi_{j} \in V_{i+k-1}.$$
Since also $kz^{k-1}\phi_{i}$ is also in $V_{i+k-1}$ it follows that $\mathcal Pz^{k}\phi_{i}$ is in $V_{i+k-1}$. 
Therefore $\mathcal P V_{i+k} \subseteq V_{i+k-1}$ and $(V_{1},\cdots,V_{n})$ is a solution to the quiver. Furthermore, one sees that if $-f$ is an exponential factor of $\nabla_{\textrm{III}}^{-}$ then so is 
$$\zeta_{n}^{i}(-f)(\zeta_{n}^{i}z)  \sim r_{0} \cdot \zeta_{n}^{i(D+1)} z^{D}$$
for some $r_{0} \in \CC^{\times}$ independent of $i$. Since we assume that $\textrm{gcd}(D+1,n)=1$ it follows that $\nabla_{\textrm{III}}^{-}$ has $n$ distinct exponential factors and one obtains $n$ distinct solutions to the string quiver with the same $B$. Looking for example at the leading order terms of the exponential factors one sees that in fact one obtains $n$ distinct gauge equivalence classes of solutions. The theorem follows.
\end{proof}
As an application one can calculate the $\CC$-dimension of the space $\mathcal S_{n,d}$ parametrizing solutions in $\mathcal S_{n}$ with the additional constraint $\textrm{deg }f \le d$.

\subsection{Classical limit}
One might expect a relation between a suitably defined classical limit of a D-module describing a quantum field theory partition function and a suitable spectral curve. We illustrate this in the context of string quivers.
To define the classical limit of a string quiver we replace 
$\mathcal P=\partial_{z^{q}}+f(z)$ by 
$\mathcal P_{\hbar}= \hbar \partial_{z^{q}} +f(z).$
Instead of connections on the punctured disc one then obtains $\h$-connections: A $\CC(\!(t)\!)$-vector space $V$ with $\CC$-linear endomorphism $\nabla$ of $V$ such that
$$\nabla(fv) = \h (\partial_{t} f) v + f \nabla(v)$$
for all $f \in \CC(\!(t)\!)$ and all $v \in V$. The classical limit of an $\h$-connection $\h \partial_{t} + M$ is the algebraic curve defined by the vanishing of the characteristic polynomial:
$$\textrm{char}(M |_{\h=0},Y)=0$$
where $Y$ is an indeterminate. Let $n$ denote the degree of the string quiver. Under the gauge transformation $g \in \mathfrak g \mathfrak l_{n} \CC(\!(t)\!)$ the matrix $M$ changes by
$M \mapsto g^{-1}M g  + \h g^{-1} \partial_{t}(g)$. 
Hence in the $\h \rightarrow 0$ limit, the characteristic polynomial of the matrix describing the $\h$-connection is well defined on the gauge equivalence class.

Suppose for example $q,n,\textrm{deg }f$ are chosen such that $\textrm{gcd}(q+\textrm{deg }f,nq)=1$ and such that one can show the following: If  $\zeta_{nq}$ is a primitive $nq$'th root of unity one has
$$ \nabla_{\textrm{II}}  \cong \bigoplus_{i=0}^{n-1}  [q]_{*}\left (\CC(\!(t)\!),\partial_{u} +\zeta_{nq}^{iq} f(\zeta_{nq}^{i}u) \right )$$
and each direct summand is irreducible. We showed for example that if $q=1$ this is the case. Another case is if $n=1$ and we now focus on this situation. When the corresponding calculations are redone incorporating the above defined $\h$-dependence one sees that there is a basis in which the $\h$-connection is of the form
$$\h \partial_{u} + \begin{bmatrix}
\ddots  & && \\
&f(\zeta_{q}^{i} u) &&\\
&&\ddots &\\
\end{bmatrix}$$
where $\zeta_{q}$ is a primitive $q$'th root of unity and $1\le i \le q$. Hence, the classical limit of the connection is given by the curve
$$\prod_{0\le i < q} \left (Y-  f(\zeta_{q}^{i} u) \right )=0.$$
This describes for example the classical limit of the quiver of the $(q,p)$ minimal model coupled to gravity, as described in Section \ref{introduction}: One has $n=1$ and $f=u^{p}$. For $X=u^{q}$ one obtains the classical limit
$$Y^{q}=X^{p}.$$

\subsection{Virasoro constraints}
\label{Virasoro-section}
Due to the relation between stabilization conditions on points in the Sato Grassmannian and relevant Virasoro constraints for the associated KP $\tau$-functions one can easily write down such Virasoro constraints for string quivers. A subtlety arises: Do relevant operators annihilate the tau functions or are there non-zero eigenvalues? Varying $n$ and $q$ turns out to have non-trivial effects. Define
$$
\textrm{J}_{m}  = 
\begin{cases}
 t_{m} \;\;\;  \;\;\;  \;\;\; \; \textrm{ if  $m>0$ } \\
 - m\partial_{t_{-m}}  \;\;  \textrm{ if } m <0
\end{cases}
$$
and for $m\in \ZZ^{\ge 0}$ define
$$\textrm{L}_{m} = \frac{1}{2} \sum_{-\infty<i<2m} \textrm{J}_{i}\textrm{J}_{2m-i}+\frac{1}{16} \delta_{m,0}.$$
We give an example of Virasoro constraints for quivers generalizing the situation of the symmetric unitary matrix model.
\begin{example}Consider 
$(n,q)=(n,1)$ with $n$ even.
Let $(V_{1},\cdots,V_{n})$ be a solution to $\textrm{\emph{String}}(n,1,f)$ and let $\tau_{i}$ denote the \emph{KP} $\tau$-function associated to $V_{i}$. Then $\tau_{i}$ is a $\tau$-function of the $n$-reduced \emph{KP} hierarchy. For $f =\sum_{k\ge 0} r_{2k} z^{2k}$ let
$$\textrm{\emph{L}}_{m}' := \textrm{\emph{L}}_{m} + \sum_{k \ge 0} r_{2k} \textrm{\emph{J}}_{2m+ 2k +1}.$$ 
Then for all $j \ge 1$ 
$$\textrm{\emph{L}}_{nj/2}' \tau_{i} = 0$$ 
and there are constants $\lambda_{1},\cdots,\lambda_{n}$ such that
$$\textrm{\emph{L}}_{0}' \tau_{i} = \lambda_{i} \tau_{i}.$$
\end{example}
To prove the above Virasoro constraints note that from the definition of a string quiver it follows that for each $i$ one has
$z^{n}V_{i} \subseteq V_{i}$ and
this is known to imply that $\tau_{i}$ is indeed a $\tau$-function of the $n$-reduced $\textrm{KP}$ hierarchy. For a point $V$ of $Gr$ let $\tau_{V}$ denote the corresponding KP $\tau$-function. It is known, see for example \cite{ABS}, that 
$$z^{2m+1}\left (\partial_{z}  + \sum_{k \ge 0} r_{2k}z^{2k} \right ) V \subseteq V  \;\;$$  if and only if  
$\textrm{L}_{m}' \tau_{V} = \lambda \tau_{V}$
for some $\lambda \in \CC$. Now note that each $V_{i}$ satisfies for all $j \ge 0$
$$z^{jn+1} \mathcal P  V_{i} =z^{jn+1}(\partial_{z}+ \sum_{k\ge 0}r_{2k}z^{2k})V_{i} \subseteq V_{i}.$$ 
It follows that there are constants $\lambda_{j,i}$ such that
$$\textrm{L}_{nj/2}' \tau_{i} = \lambda_{j,i} \tau_{i}.$$ 
For $j\ge 1$ one has 
$$\textrm{L}'_{nj/2} = \frac{2[\textrm{L}_{nj/2}',\textrm{L}_{0}']}{nj}$$ 
and it follows that 
$$\textrm{L}_{nj/2}' \; \tau_{i} = 0$$
for all $j \ge 1$. One sees that the $\textrm{L}_{0}'$ action on the $\tau$-functions is more subtle. In the case relevant for the unitary matrix model one has $n=2$ and the comparison of $\lambda_{0,1}$ and $\lambda_{0,2}$ is discussed in \cite{HMNP}.

\section{Non string quiver example}
\label{non-string-section}
In this section we consider a type of quiver slightly different than a string quiver. We will show that the relation between $\nabla_{\textrm{I}}$ and $\nabla_{\textrm{III}}^{+}$ is slightly different than for string quivers. For simplicity we set $q=1$. Suppose
$$\sigma_{\mathcal Q}(i)=i+1$$
$$\sigma_{\mathcal P}(i)= i + k$$
where everything is considered mod $n$ and $k \not \equiv -1 \mod n$. 

Suppose $(V_{1},\cdots,V_{n})$ is a solution to the quiver. Let $\phi_{i,j}$ be the preimage in $V_{i}$ of $z^{j}$ under the projection map to $\CC[z]$. Then
$$\mathcal Pz \phi_{i,j} - z\mathcal P \phi_{i,j} = \phi_{i,j}$$ 
and since the left hand side is in $V_{i+k+1}$ it follows that $\phi_{i,j}$ is in $V_{i+k+1}$ and hence
$$V_{i} = V_{i + k+1}$$
for all $i$. Suppose now that $\textrm{gcd}(k+1,n)=1$, therefore it follows
$$V_{1}= \cdots = V_{n}.$$
Furthermore, up to gauge equivalence, these points all equal $\CC[z]$. The matrix $B$ can be defined as before. Consider for example the case $n=3$ and $\sigma_{\mathcal P}=\sigma_{\mathcal Q}=(1 2 3)$ and $f=z$. Then
$$B= \begin{bmatrix}
z&0&0\\
0&z&0\\
0& 0 &z
\end{bmatrix}.$$
and
$$\nabla_{\textrm{III}}^{+} \cong  (\CC(\!(z)\!),\partial_{z} +  z)^{3}\cong \nabla_{\textrm{I}}^{3}.$$

$$$$

\textbf{Acknowledgments}: 

It is a great pleasure to thank Maarten Bergvelt, Philippe di Francesco, Thomas Nevins, Albert Schwarz for helpful exchanges.

\Addresses

\end{document}